\documentclass[journal]{IEEEtran}

\usepackage[left=1.5cm, right=1.5cm, top=1.5cm, bottom=1.5cm]{geometry}
\usepackage{amsmath}
\usepackage{graphicx} 
\usepackage{dblfloatfix}  
\usepackage{multicol}
\usepackage{array}
\usepackage{mathrsfs}
\usepackage{multirow}
\usepackage{xcolor}
\usepackage{comment}
\usepackage[export]{adjustbox}
\usepackage{subcaption}
\usepackage{enumitem}
\usepackage{verbatim}
\usepackage{algpseudocode}
\usepackage[font=small]{caption}
\usepackage[ruled]{algorithm2e}
\algnewcommand{\IfThenElse}[3]{
  \State \algorithmicif\ #1\ \algorithmicthen\ #2\ \algorithmicelse\ #3}
  
\usepackage{lipsum}
\usepackage{titlesec}

\newcommand{\alphalphval}[1]{.\arabic{equation}}
\newcolumntype{P}[1]{>{\centering\arraybackslash}p{#1}}

\AtBeginDocument{%
  \AtBeginEnvironment{subequations}{%
  }
}

\usepackage{amssymb}

\usepackage{accents}

\usepackage{booktabs}
\usepackage{comment}
\usepackage[utf8]{inputenc}
\usepackage{nomencl}
\makenomenclature
\usepackage{etoolbox}
\renewcommand\nomgroup[1]{%
    \item[\bfseries
    \ifstrequal{#1}{A}{Indices and Sets}{%
    \ifstrequal{#1}{B}{Indices}{%
    \ifstrequal{#1}{C}{Parameters}{%
    \ifstrequal{#1}{D}{Variables}{%
    \ifstrequal{#1}{E}{Abbreviations}{}}}}}%
    ]}
\makeindex

\makenomenclature

\begin{document}

\title{\huge Identifying Operation Equilibrium in Integrated Electricity, Natural Gas, and Carbon-Emission Markets}

\author{Yijie~Yang,
		Jian~Shi,~\IEEEmembership{Senior Member,~IEEE,}
		Dan~Wang$^\dag$,~\IEEEmembership{Senior Member,~IEEE,}
		Chenye~Wu,~\IEEEmembership{Member,~IEEE,}
		and Zhu~Han, ~\IEEEmembership{Fellow,~IEEE,}
		\vspace{-5mm}
	}
	
\maketitle

	
\begin{abstract}
	
 The decarbonization of the power sector plays a pivotal role in economy-wide decarbonization to set the world on track to limit warming to 1.5°C by 2050. Carbon emission markets can play a significant role in this transition by putting a price on carbon and giving electricity producers an incentive to reduce their emissions. In this paper, we study the operational equilibrium of an integrated regional/jurisdictional energy system that comprises an electricity market, a natural gas market, and a carbon emission market. We first propose the novel role of a regional carbon market operator (CMO). Different than the conventional cap-and-trade carbon trading mechanism, the proposed CMO manages carbon allowance trading in a centralized jurisdictional carbon emission market. We then develop the formulation to identify the operational equilibrium of the electricity, natural gas, and carbon emission markets based on their interactions. The proposed equilibrium can be obtained by solving the Karush-Kuhn-Tucker (KKT) conditions of all three operational models simultaneously. Simulation results demonstrate that the proposed approach is more flexible, consistent, and effective in mitigating carbon emissions compared with the cap-and-trade mechanism.
  
\end{abstract}
	
	
\vspace{-5mm}	
	

\mbox{}
\nomenclature[A]{$\mathbb{E}(i)$}{set of buses connected directly to bus $i$}
\nomenclature[A]{$\mathbb{G}(m)$}{set of nodes connected directly to node $m$}
\nomenclature[A]{$i,j$}{indices of electric buses in set $\mathbb{B}$}
\nomenclature[A]{$m,n$}{indices of natural gas nodes in set $\mathbb{N}$}
\nomenclature[A]{$t$}{time index in set $T$ }
\nomenclature[A]{$t_c$}{time index in set $T_c$ }
\nomenclature[A]{$v/V$}{index/set of generating units}
\nomenclature[A]{$w/W$}{index/set of natural gas suppliers}
\nomenclature[A]{$\Psi_i^{OG}$}{set of non-gas-fired units connected to bus $i$}
\nomenclature[A]{$\Psi_i^{WG}$}{set of wind power generating units connected to bus $i$}
\nomenclature[A]{$\Psi_i^{NG}$}{set of gas-fired generating units connected to bus $i$}
\nomenclature[A]{$\Theta_m^S$}{set of natural gas suppliers connected to node $m$}
\nomenclature[A]{$r/R$}{index/set of carbon allowance suppliers}
\nomenclature[A]{$o/O$}{index/set of carbon allowance demands}
\nomenclature[A]{$\Omega_G$}{set of natural gas-fired generating units}
\nomenclature[A]{$\Omega_R$}{set of non-natural gas-fired generating units}

\nomenclature[C]{$C^E,C^G,C^N$}{value of lost electric load [\$/p.u.]/lost natural gas load [\$/M$\rm m^3$]/unmet carbon demand [\$/ton]}

\nomenclature[C]{$C_{G,v}$}{variable production cost of non-natural gas-fired generating unit $v$ [\$/p.u.]}

\nomenclature[C]{$C_{O,v}$}{non-fuel variable operation and maintenance cost of natural gas-fired unit $v$ [\$/p.u.]}

\nomenclature[C]{$C_{M,v}$}{marginal cost of non-gas-fired unit $v$ [\$/p.u.]}

\nomenclature[C]{$C_{S,w}$}{variable production cost of natural gas supplier $w$
[\$/M$\rm m^3$]}

\nomenclature[C]{$C_{K,r}$}{carbon allowance cost for carbon allowance supplier $r$
[\$/M$\rm m^3$]}

\nomenclature[C]{$F_{S,w}^{max},F_{S,w}^{min}$}{maximum/minimum natural gas supply from supplier $w$ [M$\rm m^3$]}
\nomenclature[C]{$P_{G,v}^{max},P_{G,v}^{min}$}{maximum/minimum power output of generating unit $v$ [p.u.]}
\nomenclature[C]{$\widetilde{P}_{G,v,t}$}{predicted output of wind generating unit $v$ in time $t$ [p.u.]}
\nomenclature[C]{$P^{ramp}_{G,v}$}{ramping limit of generating unit $v$ [p.u.]}
\nomenclature[C]{$P^{max}_{i,j}$}{capacity of the line connecting buses $i$ and $j$ [p.u.]}
\nomenclature[C]{$F^{max}_{m,n}$}{capacity of the line connecting nodes $m$ and $n$ [M$\rm m^3$]}

\nomenclature[C]{$\zeta_v$}{heat rate of natural gas-fired unit $v$ [M$\rm m^{3}$/p.u.]}
\nomenclature[C]{$\eta_v$}{carbon emission rate of unit $v$ [ton/p.u.]}
\nomenclature[C]{$b_{i,j}$}{susceptance of line connecting buses $i$ and $j$ [p.u.]}
\nomenclature[C]{$Q_{L,o,t_c}$}{carbon allowance demands in hour $t$ [ton]}
\nomenclature[C]{$Q_{C,r,t_c}^{max}$}{maximum carbon allowance supply from supplier $r$ [ton]}

\nomenclature[D]{$\theta_{i,t}$}{phase angle of bus $i$ in hour $t$ [rad]}

\nomenclature[D]{$F_{S,w,t}$}{natural gas supplied in hour $t$ by supplier $w$ [M$\rm m^3$]}

\nomenclature[C]{$P_{L,i,t}$}{electric demand at bus $i$ in hour $t$ [p.u.]}
\nomenclature[C]{$F_{L,m,t}$}{non-generation-related natural gas demand at node $m$ in hour $t$ [M$\rm m^3$]}
\nomenclature[D]{$F_{L,m,t}^D$}{non-generation-related natural gas demand that is served in hour $t$ at node $m$ [M$\rm m^3$]}
\nomenclature[D]{$F_{m,n,t}$}{natural gas flow through the pipeline connecting nodes $m$ and $n$ in hour $t$ [M$\rm m^3$]}
\nomenclature[D]{$P_{G,v,t}$}{active power output from generating unit $v$ in hour $t$
[p.u.]}
\nomenclature[D]{$P_{L,i,t}^D$}{electric demand that is served at bus $i$ in hour $t$ [p.u.]}
\nomenclature[D]{$Q_{L,o,t_c}^D$}{carbon allowance demand $o$ that is served in hour $t_c$ [p.u.]}
\nomenclature[D]{$Q_{C,r,t_c}$}{carbon allowance supplied in hour $t$ by supplier $r$ [ton]}
\nomenclature[D]{$p_{co_2,t}$}{carbon price in hour $t$ [\$/ton]}

\nomenclature[E]{CEM}{Carbon emission market}
\nomenclature[E]{CMO}{Carbon market operator}
\nomenclature[E]{ETS}{Emission trading system}

\printnomenclature[2cm]

\section{Introduction}

Power system is a major carbon emission producer: approximately 60\% of the United States’ electricity comes from fossil fuel combustion, making up 25\% of the total U.S. greenhouse gas (GHG) emissions \cite{wheregreenhouse}. In addition, with the rapid electrification of the building, industrial, and transportation sectors, the demand for power is projected to surge in the next decade . As an example, electrification is projected to increase annual electricity use in ISO New England by 14\% over the next decade, according to \cite{ISONewEngland}. Looking forward, deep and prompt decarbonization of the power sector is critical across many of the envisioned pathways to a decarbonized economy \cite{IPCC2021}.

With rising generation from natural gas, wind, and solar, the U.S. power sector has been decarbonizing at an average rate of 3\% per year since 2008 \cite{Decarbonizing}. From 2010 to 2019, coal-fired generation declined by 48\%, and natural gas generation increased by 58\% \cite{coalfiredelectricity} due to the breakthrough in natural gas extraction techniques (e.g., hydraulic fracturing) and the superior characteristics of the natural-gas fired generation. The total global renewable generation has also reached a record of 795 million MWh in 2021, accounting for about 20.1\% of the electricity generation \cite{RenewableGeneration}. Despite the abovementioned progress, it is pointed out in \cite{2035r} that the current emission trends are not on track to meet the urgent climate goal envisioned by the global leaders (e.g., a national 80\% clean electricity share by 2030 for the U.S. \cite{FACTSHEET}). The electric power system needs to drastically accelerate its current trajectory toward carbon neutrality. Given the urgency of such a transition, it is evident that a multi-faceted approach is required, including technical measures such as the continued substitution of zero-/low-emission power sources, improvements in end-use efficiency, large-scale deployment of storage as a grid flexibility resource, and use of carbon capture, utilization, and storage (CCUS). On the other hand, innovations in policy and regulations are equally important to stimulate near-term actions from stakeholders, address system-wide impacts of decarbonization, and facilitate the net-zero energy transition. 

Currently, two main emission policy instruments existed to mitigate carbon emissions and facilitate the green energy transition \cite{Carbon_handbook}: 1) Carbon emission taxes/credits, and 2) Cap-and-trade carbon emission market (CEM). Carbon emission tax takes the form of a predetermined fixed tax rate, which is applied to energy-generating companies that produce carbon emissions. When facing a carbon tax, energy producers will seek to minimize the cost by deploying emission reduction strategies and substituting away from carbon-intensive technologies and operations. On the other hand, in a CEM, an allowable amount of emissions (i.e., emission cap) is imposed by the policy-maker on the energy system to explicitly limit the overall emissions. This allowance, typically measured by tonnes of CO\textsubscript{2}, is then assigned to energy producers who make their energy generation decisions accordingly and ensure that their emissions do not exceed the emission allowance they hold. CEM participants can also trade their unused carbon allowances in a market environment. Compared with carbon taxes, CEM focuses on policy certainty by explicitly limiting the overall emissions cap, while providing robust price incentives for reducing fossil-fuel-based energy consumption and investing in clean low-carbon technologies. Due to its attractiveness, as of 2021, CEMs have been implemented or are under development in 38 countries across four continents \cite{Carbon_handbook}, with European Union (EU) emission trading system (ETS) being the world's largest CEM, covering roughly 41\% of the total EU GHG emissions and accounted for 84\% of the value of the total global CEMs. During the past decade, CEMs have been proven effective. For instance, since its establishment in 2005, EU ETS has resulted in a 42.8\% emission reduction to date \cite{EU_ETS}.

However, making the CEM a part of the existing energy ecosystem is not straightforward. One of the most critical challenges is that CEM and other emission regulation schemes should not disrupt the economics of electricity, as an essential commodity linked to social welfare. Currently, given the dominant role natural gas is playing in electricity generation (e.g., natural gas was the largest source, roughly 38\%, of the U.S. electricity generation in 2021 \cite{Electricityexplained}), the price of natural gas impacts the generation cost of natural gas-fired generators (e.g., steam turbines and gas turbines). Therefore, power generation companies (Gencos), as profit-seeking entities, need to participate in both the electricity market and natural gas market (i.e., the fuel market) to maximize their profits. To meet the decarbonization goals, Gencos, as emitters, also need to participate in the CEM, as carbon pricing would ultimately affect the decision-making process of the Gencos. Higher carbon prices would provide a stronger market signal to reward emission abatement measures and renewable generation from Gencos. However, it may increase the cost of the generation, resulting in higher electricity prices. On the other hand, lower carbon prices may fail to incentivize Gencos and lead to the failure of achieving decarbonization objectives. Therefore, it is evident that electricity markets, natural gas markets, and carbon emission markets are becoming ever-increasing intertwined. While these three markets are operated independently by different system operators, their operations are all interrelated, and therefore need to be coordinated to achieve efficient and effective operation while meeting the decarbonization goals.

In the literature, the interdependencies between carbon emissions and the energy markets have been studied from two perspectives: 1) empirical energy-economic approaches, and 2) optimization-based approaches. More specifically, for energy-economic techniques, a standard econometric approach was developed in \cite{2020An} to study the interactions between electricity, carbon, and fossil fuel prices. The findings support the existence of a long-term co-integration between the price of carbon and those of coal, diesel, and liquefied natural gas (LNG). A carbon-electricity linkage model was proposed in \cite{2020Linking} based on a trading mechanism that involves PV power plants and links carbon-electricity markets. Simulation results show that the carbon-electricity linkage trading mechanism can offset the negative impact of PV levelized cost of electricity (LCOE) rise brought by the subsidy decline on the PV industry. The dynamic multi-scale interactions between carbon and electricity markets during EU ETS phases I (2005–2007), II (2008–2012), and III (2013–2016) were explored in \cite{2017Dynamic} by introducing the bivariate empirical mode decomposition (BEMD), linear and nonlinear Granger causality tests. The findings demonstrated that the CEM's influence on the electricity market is limited, indicating the need to raise the price of carbon to improve its efficiency. Despite their policy relevance, these works are largely empirical and concentrated on gaining economic and policy insights through the study of historical data and simulation experiments. There is a lack of in-depth modeling of the CEM operation and how a CEM interacts with other energy markets.

To address this issue, optimization models have been explored in the literature. For instance, a multi-objective approach was developed in \cite{2008Profit} to handle the problem of conflicting profit and emission objectives. It yielded a trade-off curve between profit and pollution that will help decision-makers regarding the trading of emission allowances. Both the cap-and-trade model and carbon tax policies were investigated in \cite{he2012cap} to find the optimal solution to the generation expansion investment equilibrium problem. Under cap-and-trade, an overall emission cap was imposed on the energy system, while the electricity generation was taxed on a \$/MW basis under the carbon taxes policy. A robust environmental-economic dispatch (EED) method was proposed in \cite{2016Robust} that jointly optimizes energy and reserve schedules in the upcoming dispatch period. The Nash bargaining criterion is adopted to determine a fair trade-off between the generation cost and the carbon emission in the absence of a clear carbon tax or emission cap. A conjectural-variations equilibrium model was proposed in \cite{chen2021conjectural} to study the equilibria reached in electricity, natural gas, and carbon emission markets. A simple cap-and-trade mechanism was used to model the carbon-emission market. While these works incorporated the CEM model into their optimization problem, the operation of the CEM was modeled as a simple emission cap that Gencos must follow. The underlying trading mechanism, pricing scheme, and market environment of the CEM have yet to be fully explored in the literature. 

This paper aims to bridge this important gap by establishing an integrated regional/jurisdictional energy system model that is composed of an electricity market, a natural gas market, and a CEM. We first propose the operational model of a regional carbon operator, who operates and manages centralized trading in a jurisdictional CEM. We then characterize the operational equilibrium of the three aforementioned markets by deriving the Karush-Kuhn-Tucker (KKT) conditions and transform the resulting mathematical programming with equilibrium constraints (MPEC) problem into a mixed-integer linear programming (MILP) problem that can be solved by commercial solvers. Lastly, we examine the performance of the proposed approach in comparison to the conventional cap-and-trade mechanism.

The contributions of this paper are summarized as follows:
\begin{enumerate}
 \item {}We establish the novel role of a regional carbon market operator. The proposed operator manages the centralized trading of carbon allowance in a jurisdictional/regional CEM. To the best of our knowledge, this paper is the first of its kind.
 
 \item { We develop an MPEC approach to identify the operational equilibrium of the integrated electricity, natural gas, and carbon emission market. The proposed approach pinpoints market operation that are simultaneously optimal in all three markets, which provides a quantitative way to balance the near-term social welfare and the urgency of meeting the decarbonization goal. }
 
 \item { We systematically show the strength of the proposed approach compared with the conventional cap-and-trade mechanism heavily adopted by current CEM designs. Simulation results presented in this paper suggest that the proposed approach is more flexible, consistent, and effective in accelerating the deep decarbonization of the electric power system.}

 \end{enumerate}

The remainder of this paper is organized as follows. Section II presents the proposed regional CEM and the proposed role of CMO. Section III presents the operational models of the electricity and natural gas market as well as the identification of their operational equilibrium. Section IV illustrates the solution methodology to derive such an equilibrium. The simulation-based case studies are carried out in Section IV, and finally, the conclusions are drawn in Section V.

\section{Regional CEM Model}

In this section, we perform a brief review of the basic operational principles involved in the cap-and-trade CEM. We then review the current cap-and-trade model that has been adopted in the literature. Finally, we describe the proposed regional CEM model in terms of its key mechanisms and formulation.

\subsection{Basics of Cap-and-trade CEM}

When it comes to CEM design, the EU ETS has been commonly used as the benchmark model. The EU ETS is a multi-country, multi-sector environmental law mandated by the EU Commission. The EU ETS is a “cap-and-trade" system, which works by capping overall GHG emissions of more than 11,000 power stations and industrial plants across the EU in the form of available allowances. The cap level is determined by the EU legislative branches and is designed to decrease annually in a linear fashion. Each year, a proportion of the allowances are given to certain participants for free, while the rest are allocated through auctions. At the end of a year, all the participants must return an allowance for their emissions during that year. If a participant has insufficient allowances, then it must either take measures to reduce its emissions or purchase more allowances on the market from other market participants. Compliance is ensured through penalties (e.g., €100/CO2 tonne) and other enforcement structures \cite{EU_ETS}. 



In terms of carbon trading, emission allowance transactions are conducted directly between buyers and sellers, commonly referred to as “Over-the-Counter” (OTC), through organized exchanges. Market participants can also buy or sell from intermediaries, such as banks and specialist traders. Overall, the current carbon exchange mechanism of the EU ETS can be represented as a bilateral market. While the market needs to be supervised by the CMO, it is primarily relying on negotiable agreements among the market participants. In other words, the carbon allowance holders and buyers need to set up contracts on the amount, price, and delivery format of the carbon allowances, independent of the CMO. 


\subsection{Current Cap-and-trade CEM Model}

In the literature, the cap-and-trade CEM is commonly modeled according to the cap-and-trade mechanism described in the previous section \cite{zhao2010long}\cite{chen2021conjectural}\cite{he2012cap}. More specifically, the following complementarity condition has been used to derive the carbon price in the CEM:

\begin{align}\label{c1}
&0 \leq \left(Cap-\sum_{v \in V, t \in T }{\eta_{v}P_{G,v,t}}\right)\perp p_{co_2} \geq 0  
\end{align}

\noindent where $Cap$ is the cap of the carbon allowance assigned to the CEM, $\eta_{v}$ is the emission rate of generation unit $v$, $P_{G,v,t}$ denotes the generation output of $v$ at a particular hour $t$. The carbon price $p_{co_2}$ is the shadow price of the total carbon limit constraint. Note that the form $0 \leq a \perp b \geq 0 $
is equivalent to $a \geq 0$, $b \geq 0$ and $ab=0$.

Constraint (\ref{c1}) indicates that if the total carbon emission is equal to the total cap allocated, the carbon price is positive; otherwise, it is zero.
\subsection{Proposed Regional CEM Model}
While the existing practices provide insights into the design and implementation of CEMs, it is evident that there is no universal solution for such a challenging problem. Policymakers and regulators of jurisdictions need to decide on the major design features, as well as the key processes of their carbon trading mechanisms.

In this paper, we consider a regional CEM. Different from larger inter-country carbon trading platforms, a regional CEM can be more scalable, equitable, and accessible for small carbon consumers and allowance holders who prefer to trade locally or have trouble getting direct access to the larger market. Dividing a large geographical area into multiple regional carbon jurisdictions and assigning regional carbon operators also enables the decision-makers to take into account the economic developments and reduction potentials that are unique to different regions. As the managing organization of the regional CEM, the regional CMO can work with the jurisdictional policy-makers to administer the carbon emission allowance market, set carbon prices, manage the carbon exchange among local market participants, and interact with larger, interconnected CEM operators, such as EU-ETS, as a carbon allowance aggregator. The regional CEM can also be a voluntary market in which the trade of carbon credits and/or allowance is on a voluntary basis. In either case, the goal of the CMO is to stay on track with the goal of decarbonization while facilitating the cost-effective operation of the CEM and providing appropriate carbon pricing for other energy markets within the region. 

\begin{figure}[h]
\centering
\includegraphics[scale=0.7]{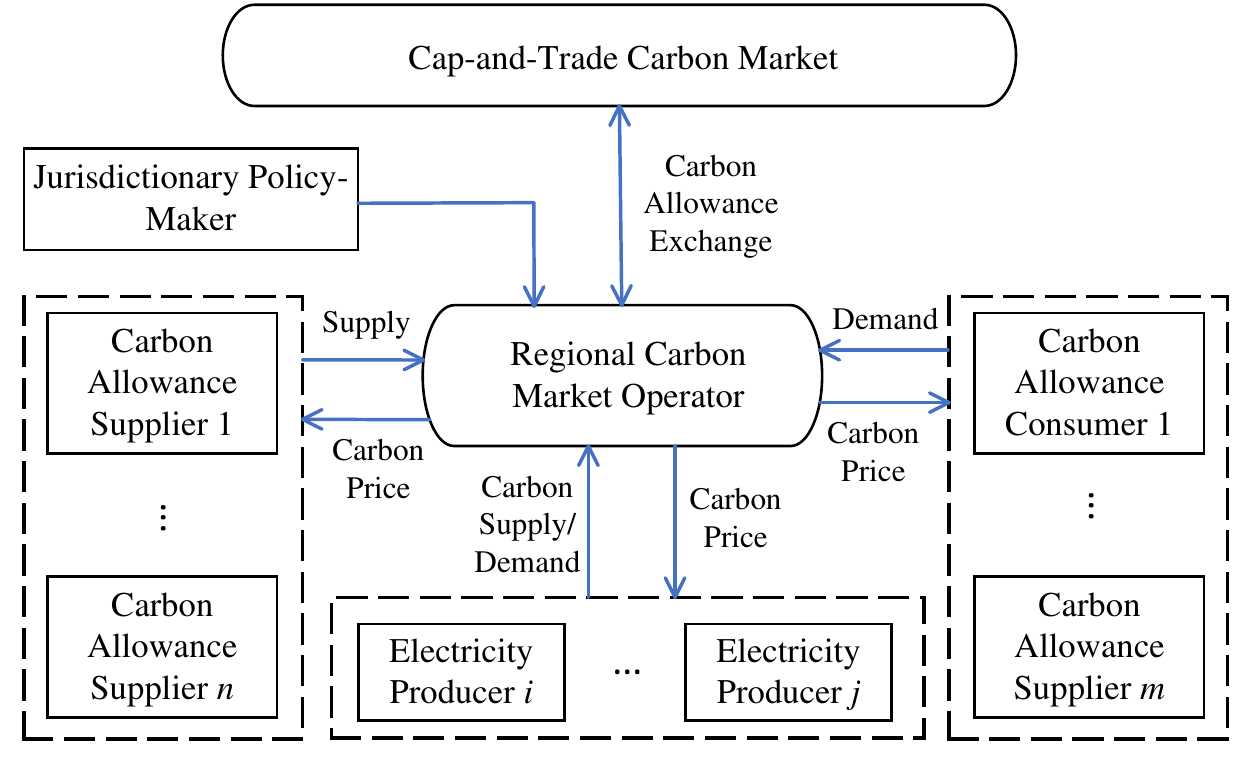}
\caption{Proposed regional CEM and CMO.}
\label{RegionalCMP}
\end{figure}

Based on the previous discussion, it is evident that the regional CMO is capable of playing a bigger role than a simple market facilitator. In this paper, we consider a PoolCo (Pool) model which provides a centralized marketplace for carbon allowance buyers and sellers. Similar to other forms of energy markets, market participants, including Gencos and other carbon allowance holders and consumers in the pool, submit bids and offers to the pool for the amounts of carbon allowances they are willing to buy/sell. The CMO, as the managing entity of the CEM, has full control of the carbon allowance exchanges in the system and produces a market clearing price for carbon allowances, which gives participants the market signal to stimulate emission mitigation actions as shown in Fig. \ref{RegionalCMP}. More specifically, we propose a CEM operational problem as follows:

\begin{align} 
\label{c2}
\min _{\Xi_{\mathrm{E}}} \sum_{t \in T_c} & \left[ \sum_{ r \in R} C_{K, r}Q_{C,r,t_c}
+ \sum_{o \in \mathrm{O}}C^{N} \cdot(Q_{L,o,t_c}- Q_{L,o, t_c}^{D})\right]
\end{align}

\begin{align}
\text{s.t.}
&\sum_{r \in R} Q_{C, r, t_c}=\sum_{t \in {\{t_c\}}}\sum_{v \in \Theta_{i}^{\mathrm{G}}} \eta_{v,t}P_{G,v, t}+\sum_{o \in O}Q_{L,o,t_c}^D :p_{co_2,t_c}; \notag \\
&\qquad  \qquad \qquad \qquad \qquad  \qquad\qquad \forall i \in \mathcal{I}, \forall t \in T_c  \label{c3}\\
&0 \leq Q_{L,o, t_c}^{\mathrm{D}} \leq Q_{L,o,t_c}: (\nu_{1,m,t}^{min},\nu_{1,m,t_c}^{max});\forall t_c \in T_c, \forall o \in O \label{c4}\\
&0 \leq Q_{C,r, t_c} \leq Q_{C,r,t_c}^{max}: (\nu_{2,r,t_c}^{min},\nu_{2,r,t_c}^{max});\forall t_c \in T_c, \forall r \in R \label{c5}
\end{align}
where $Q_{C, r, t_c}$ denotes the carbon allowance supplied in time period $t_c$ by supplier $r$, $Q_{L, o, t_c}^D$ is the carbon allowance demand $o$ that is served in time period $t_c$. Since CEMs may have a longer time clearing scalar than electricity markets, $\{t_c\}$ denotes the time set consisting of $t$ which belongs to the longer time period $t_c$.
The primal optimization variables of the problem are included in the set $\Xi_{\mathrm{C}}^P=\{Q_{C, r, t_c}, Q_{L,o,t_c}^{D}\}$, while the set of dual variables is $\Xi_{\mathrm{C}}^D=\left\{ p_{co_2,t_c}, \nu_{1,r,t_c}^{min}, \nu_{1,r,t_c}^{max},\nu_{2,o,t_c}^{min}, \nu_{2,o,t_c}^{max} \right\}$.

The objective function (\ref{c2}) computes the overall operation cost of the CEM. The first term in the objective function represents the cost of carbon allowance supply, while the second term computes the cost associated with the unmet carbon consumption demand. Constraint (\ref{c3}) imposes the balance of carbon allowance consumption and supply. The dual variable, $p_{co_2,t_c}$, that is associated with (\ref{c2}) represents the carbon price in time period $t_c$. Constraint (\ref{c4}) bounds the carbon allowance demands served, which ensures the progress of decarbonization. Constraint (\ref{c5}) limits carbon allowances held by different carbon allowance suppliers. 

It can be observed from the model above that the goal of the CMO in the proposed market is to minimize the cost of CEM operation, balancing the carbon allowance supply and demand while meeting the emission mitigation goals. The market dynamics described above would drive the market clearing price of carbon allowance $p_{co_2,t_c}$ to a competitive level which is equal to the marginal carbon allowance cost of the most efficient bidders.

It is important to note that the above formulation assumes that carbon allowances can be exchanged between the electricity sector and other sectors of the economy. We also assume that the proposed model does not account for the transaction costs of allowance trading. Furthermore, it should be noted that compared with other forms of energy markets, such as the electricity market and natural gas market which typically operate on a day-ahead basis, a CEM can have a different, and commonly longer settlement time based on the current CEM design practices.





\section{Operational Equilibrium of Electricity, Natural gas, and Carbon Emission Markets}

In this section, we develop the modeling strategy for the electricity market as well as the natural gas market. We then propose a direct approach to identify the operational equilibria among these two markets and the previously defined CEM.

\subsection{Electric Power Market Operation Model}

To keep the electric power system in balance, the regional electric system operator (e.g., an independent system operator (ISO)) needs to decide the hourly production of each power plant in the day-ahead market based on the projected load profiles. Meanwhile, the ISO also needs to take into account the limits of the generating units, as well as the capacity of the transmission network, to ensure the feasibility of the dispatch decisions \cite{8762124}. 

The power system-operational problem is formulated as:

\begin{align} \label{e1}
\min _{\Xi_{\mathrm{E}}^P} \sum_{t \in T} & \left[ \sum_{m \in \mathbb{N}, v \in \Psi_{m}^{OG}}\left(C_{O, v}+\zeta_{v} \mu_{m(v),t}+\eta_{v} p_{co_2,t}\right) P_{G, v, t} \right. \notag \\
&+\sum_{m \in \mathbb{N}, v \in \Psi_{m}^{NG}}\left(C_{G, v}+\eta_{v} p_{co_2,t}\right) P_{G, v, t} \notag\\
&\left.+\sum_{i \in \mathbb{B}} C^{E} \cdot\left(P_{L, i, t}-P_{L, i, t,s}^{D}\right)\right]
\end{align}
\begin{align}
\text { s.t. } 
&\sum_{v \in \Theta_{i}^{G}} P_{G, v, t}+\sum_{v \in \Theta_{i}^{WG}}\widetilde{P}_{G,v,t}=P_{L, i, t}^{D} \notag 
\end{align}
\begin{align}
&+\sum_{j \in \mathbb{E}(i)} b_{i, j}\cdot\left(\theta_{i, t}-\theta_{j, t}\right): \lambda_{i,t};\forall i \in \mathbb{B}, \forall t \in T \label{e2} 
\end{align}
\begin{align}
&P_{G, v}^{\min } \leq P_{G, v, t} \leq P_{G, v}^{\max } :(\rho_{1,v,t}^{min},\rho_{1,v,t}^{max}) ;\notag\\
&\qquad\qquad \qquad  \qquad \qquad\qquad \forall v \in \Omega_{G} \cup \Omega_{R} ,
\forall t \in T \label{e3}\\
&-P_{G, v}^{\mathrm{ramp}} \leq P_{G, v, t}-P_{G, v, t-1} \leq P_{G, v}^{\mathrm{ramp}} :\notag\\
&\qquad\qquad \qquad   (\rho_{2,v,t}^{min},\rho_{2,v,t}^{max}); \forall v \in \Omega_{G} \cup \Omega_{R} , \forall t \in T  \label{e4}\\
&-P_{i, j}^{\max } \leq b_{i, j} \cdot\left(\theta_{i, t}-\theta_{j, t}\right) \leq P_{i, j}^{\max } :(\rho_{3,i,j,t}^{min},\rho_{3,i,j,t}^{max}) \notag\\
&  \qquad \qquad \qquad  \qquad\qquad \qquad  \forall i \in \mathbb{B} ,j \in \mathbb{E}_i, \forall t \in T \label{e5} \\
&\theta_{\mathrm{REF}, t}=0 : \rho_{4,t};\forall t \in T \label{e6}\\
&0 \leq P_{L, i, t}^{D} \leq P_{L, i, t} :(\rho_{5,i,t}^{min},\rho_{5,i,t}^{max}); \forall i \in \mathbb{B} ,\forall t \in T \label{e7}
\end{align}
where $P_{G,v,t}$ denotes the hour-$t$ active power output from generating unit $v$, $P_{L,i,t}^D$ is the electric demand that is served in hour $t$ at bus $i$, $\theta_{i,t}$ is the phase angle of bus $i$ in hour $t$, the parameter $\widetilde{P}_{G,v,t}$ is the predicted output of wind generating unit $v$ in hour $t$. The primal optimization variables of the problem are included in the set $\Xi_{\mathrm{E}}^P=\{P_{G, v, t}, P_{L, i, t}^{D}, \theta_{i, t}\}$, while the set of dual variables is $\Xi_{\mathrm{E}}^D=\left\{ \lambda_{i,t}, \rho_{1,v,t}^{min}, \rho_{1,v,t}^{max},\rho_{2,v,t}^{min}, \rho_{2,v,t}^{max} ,\rho_{3,i,j,t}^{min}, \rho_{3,i,j,t}^{max},\rho_{4,t},\rho_{5,i,t}^{min}, \rho_{5,i,t}^{max}\right\}$.

The objective function (\ref{e1}) aims to minimize the cost of the electricity market. The first term in (\ref{e1}) is the total operation cost of the power generation units, while the second term denotes load-curtailment cost. Specifically, the terms, $\zeta_{v}\mu_{v,t}$ and $\eta_{v}p_{co_2,t}$  represent the variable fuel cost of gas-fired unit $v$ and the carbon emission cost of unit $v$ in hour $t$,  respectively. $m(v)$ denotes the gas nodes where gas-fired unit $v$ is located. The power balance at each bus is represented by Constraint (\ref{e2}) where its dual variable corresponds to the locational marginal price (LMP) at each bus. The dual variable $\lambda_{i,t}$ of constraint (\ref{e2}) can be interpreted as the locational marginal price (LMP) at each bus. This is because in a convex problem, some of its dual variables can be interpreted as well-behaved prices \cite{o2005efficient}. Constraint (\ref{e3}) imposes the capacity limits of each generation unit. Constraint (\ref{e4}) imposes the ramping limits for the generation units. Constraint (\ref{e5}) enforces the transmission capacity of each power line.  Constraint (\ref{e5}) enforces the capacity limit of each transmission line where the power flow though line  $i-j$ in hour $t$ $P_{i,j,t}$ is $b_{i,j}(\theta_{i,t}-\theta_{j,t})$. Observe that   $j \in \mathbb{E}_i$ denotes the buses $j$  that are connected to bus $i$. Constraint (\ref{e6}) sets the phase angle at the reference bus. Finally, constraint (\ref{e7}) bounds the electricity demands served in the electricity system.

\subsection{Natural Gas Market Operation Model}
Similar to the electricity market, we consider a regional day-ahead natural gas market that is operated by the natural gas transmission system operator (NG-TSO). The natural gas system-operational problem is formulated as:
\begin{align}\label{g1}
\min _{\Xi_{\mathrm{G}}^P} \sum_{t \in T} & \left[ \sum_{ w \in W} C_{S, w}F_{S,w,t}
+ \sum_{m \in \mathrm{M}}C^{G} \cdot(F_{L,m,t}- F_{L,m, t}^{D})\right]
\end{align}

\begin{align}
\text { s.t. } 
& F_{L, m, t}^D+\sum_{v \in \Psi_m^{NG}}\zeta_vP_{G,v,t}=\sum_{w \in \Theta_{m}^{G}}F_{S, w, t}+\sum_{n \in \mathbb{E}(m)}F_{m,n,t} :\notag \\
&\qquad \qquad \qquad \qquad \qquad \qquad  \mu_{m,t}; \forall m \in \mathbb{N},  \forall t \in T \label{g2}
\end{align}
\begin{align}
&F_{w}^{\min } \leq F_{S,w, t} \leq F_{ w}^{\max } :(\phi_{1,w,t}^{min},\phi_{1,w,t}^{max}) ;\notag\\
&\qquad \qquad\qquad \qquad \qquad \qquad \forall v \in \Omega_{G} \cup \Omega_{R} , \forall t \in T \label{g3}
\end{align}
\begin{align}
&-F_{m,n,t}^{\max } \leq F_{m,n,t} \leq F_{m,n,t}^{\max } :(\phi_{2,m,n,t}^{min},\phi_{2,m,n,t}^{max}) \notag\\
&  \qquad \qquad \qquad  \qquad \forall m \in \mathbb{B} ,n \in \mathbb{E}_m, \forall t \in T \label{g4}\\
&0 \leq F_{L, m, t}^{D} \leq F_{L, m, t} :(\phi_{3,m,t}^{min},\phi_{3,m,t}^{max}) \quad \forall i \in \mathbb{B} ,\forall t \in T \label{g5}
\end{align}
where $F_{S,w,t}$ denotes natural gas supplied in hour $t$ by supplier $w$, $F_{L,m,t}^D$ is the natural gas demand at node $m$ that is served in hour $t$, $F_{m,n,t}$ is hour-$t$ natural gas flow through the pipeline connecting nodes $m$ and $n$. The primal optimization variables of the problem are included in the set $\Xi_{\mathrm{G}}^P=\left\{F_{S,m, t}, F_{L, m, t}^{D}, F_{m, n, t}\right\}$, while the set of dual variables is $\Xi_{\mathrm{G}}^D=\left\{ \mu_{m,t}, \phi_{1,w,t}^{min}, \phi_{1,w,t}^{max},\phi_{2,mn,t}^{min}, \phi_{2,mn,t}^{max} ,\phi_{3,m,t}^{min}, \phi_{3,m,t}^{max}\right\}$.

The objective function (\ref{g1}) is the cost of the natural gas market. The first term in (\ref{g1}) is the cost of each natural gas source, while the second term computes the natural gas load-curtailment cost. Constraint (\ref{g2}) imposes the natural gas balance on each node. The dual variable associated with (\ref{g2}), $\mu_{m,t}$, represents the natural gas locational marginal price at node $m$ in hour $t$. Constraint (\ref{g3}) limits the maximum gas flow of gas sources. Constraint (\ref{g4}) constrains the maximum gas flow on each pipeline. Constraint (\ref{g5}) bounds the natural gas demand at each node.

\subsection{Operational Equilibrium Among the Electricity, Natural Gas, and Carbon Emission Markets}

\begin{figure}[t]
\centering
\includegraphics[scale=0.6]{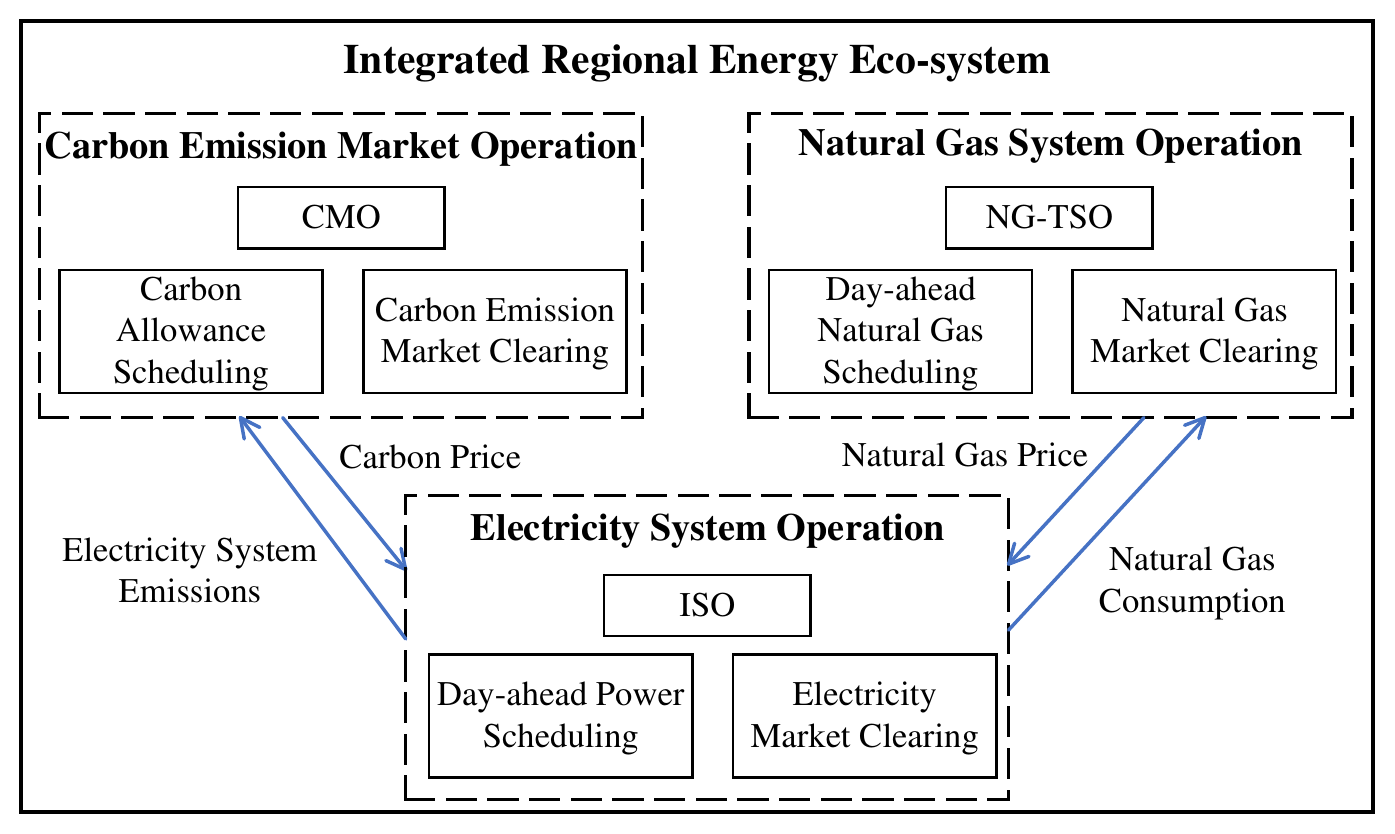}
\caption{Structure of the proposed integrated electricity, natural gas, and carbon emission markets.}
\label{fig:model}
\end{figure}

Based on the above discussion, it is evident that although the electricity, natural gas, and carbon emission markets are modeled independently, their operations are closely interdependent as depicted in Fig. \ref{fig:model}. More specifically, the interrelations among the operations of the three systems can be described as follows:

1) Between the electricity market and CEM: $P_{G,v,t}$ is a decision variable in the electricity market model. It also appears in the carbon emission balance constraint (\ref{c3}), which belongs to the CEM model. Besides, carbon price $p_{co_2,t}$, which is the dual-variable of the carbon emission balance constraint (\ref{c3}), also appears in the objective function of electricity market (\ref{e1}). Note that the two markets have different market clearing times, and the CEM clearing time $t_c$ is commonly greater than that of the electricity market $t$. Therefore, we suppose $p_{co_2,t}$ equals to $p_{co_2,t_c}$ if time $t$ is within the time period $t_c$. Namely, we have
\begin{align}
&p_{co_2,t} = p_{co_2,t_c}, \quad if \quad t \in \{t_c\}
\end{align}

2) Between electricity and natural gas markets: $P_{G,v,t}$ is a decision variable in the electricity market model, which also appears in the natural gas balance constraint (\ref{g2}) as a part of the natural gas market operation model. Besides,  natural gas LMP $\mu_{m,t}$, which is the dual-variable of the natural gas balance constraint (\ref{g3}), also appears in the objective function of the electricity market (\ref{e1}).

Note that we assume that the electric and natural gas systems have perfect price-based coordination. In this way, the natural gas prices used to dispatch the electric power system perfectly reflect the corresponding true natural gas LMPs. This is consistent with the current gas-power coordination practices. Meanwhile, the CMO is also fully aware of the true prices of the other two markets.

Based on the above discussion, we are interested in identifying the operational equilibrium of the three systems that is simultaneously optimal for all three system operators. In other words, none of the system operators can reduce the operation cost of their respective markets from deviating from such an equilibrium.

\section{SOLUTION METHODOLOGY}

As the models for the electricity, natural gas, and carbon emission markets are all linear, we can use a direct approach to find the operational equilibrium by simultaneously solving their necessary and sufficient Karush-Kuhn-Tucker (KKT) conditions. 



\subsection{Optimality Conditions for Electric Power Market Operation Model}

For the electricity system, a set of necessary and sufficient conditions for an optimum of (\ref{e1})–(\ref{e7}) is (\ref{e2})–(\ref{e7}) and:
\begin{align}
&(C_{o,v}+\eta_{o,v}\cdot p_{co_2,t})-\lambda_{i(v),t}-\rho_{1,v,t}^{min}+\rho_{1,v,t}^{max}-\rho_{2,v,t}^{min} \notag\\
&+\rho_{2,v,t}^{max}+\rho_{2,v,t+1}^{min}-\rho_{2,v,t+1}^{max}=0;\forall v \in \Omega_R, \forall t \in T;\label{de1}\\
&(C_{G,v}+\zeta_{v} \mu_{m(v),t}+\eta_{o,v}\cdot p_{co_2,t})-\lambda_{i(v),t}-\rho_{1,v,t}^{min}+\rho_{1,v,t}^{max} \notag\\
&-\rho_{2,v,t}^{min}+\rho_{2,v,t}^{max}+\rho_{2,v,t+1}^{min}-\rho_{2,v,t+1}^{max}=0;\forall v \in \Omega_G, \forall t \in T;\label{de2}
\end{align}
\begin{align}
&-C^E+\lambda_{i,t}-\rho_{5,i,t}^{min}+\rho_{5,i,t}^{max}=0;\forall i \in I, \forall t \in T;\label{de3}\\
&\sum_{j \in \mathbb{E}_i} b_{i,j}(\lambda_{i,t}-\lambda_{j,t})+\sum_{j \in \mathbb{E}_i} b_{i,j}(\rho_{3,i,j,t}^{max}-\rho_{3,j,i,t}^{max}) \notag\\
&-\sum_{j \in \mathbb{E}_i} b_{i,j}(\rho_{3,i,j,t}^{min}-\rho_{3,j,i,t}^{min})=0;\forall i \in \mathbb{B} ,i \neq REF,  \forall t \in T  \label{de4}
\end{align}
\begin{align}
&\sum_{j \in \mathbb{E}_i} b_{i,j}(\lambda_{i,t}-\lambda_{j,t})+\sum_{j \in \mathbb{E}_i} b_{i,j}(\rho_{3,i,j,t}^{max}-\rho_{3,j,i,t}^{max}) \notag\\
&-\sum_{j \in \mathbb{E}_i} b_{i,j}(\rho_{3,i,j,t}^{min}-\rho_{3,j,i,t}^{min})+\rho_{4,t}=0; i = REF ,\forall t \in T  \label{de5}\\
&P_{G, v, t}-P_{G, v}^{\min } \geq 0   \perp \rho_{1,v,t}^{min}\geq 0  ; \forall v \in V, \forall t \in T\label{de6}\\
&P_{G, v}^{\max }-P_{G, v, t} \geq 0   \perp \rho_{1,v,t}^{max}\geq 0  ;\forall v \in V, \forall t \in T\label{de7}\\
&P_{G, v, t}-P_{G, v, t-1} -P_{G, v}^{\mathrm{ramp}} \geq 0  \perp \rho_{2,v,t}^{min}\geq 0  ;\forall v \in V, \forall t \in T\label{de8}\\
&P_{G, v}^{\mathrm{ramp}}-(P_{G, v, t}-P_{G, v, t-1}) \geq 0  \perp \rho_{2,v,t}^{max}\geq 0  ; \forall v \in V, \forall t \in T\label{de9}\\
&b_{i, j} \cdot\left(\theta_{i, t}-\theta_{j, t}\right) -P_{i, j}^{\max } \geq 0  \perp \rho_{3,i,j,t}^{min}\geq 0  ; \forall v \in V, \forall t \in T\label{de10}\\
&P_{i, j}^{\max }-b_{i, j} \cdot\left(\theta_{i, t}-\theta_{j, t}\right) \geq 0  \perp \rho_{3,i,j,t}^{max}\geq 0  ; \forall i \in \mathbb{B} ,j \in \mathbb{E}_i,\notag \\
&\qquad \qquad \qquad \qquad \qquad \qquad \qquad \qquad \qquad \qquad\forall t \in T  \label{de11}\\
&  P_{L, i, t}^{D}\geq 0   \perp \rho_{5,i,t}^{min}\geq 0  ; \forall i \in I, \forall t \in T\label{de12}&\\
&  P_{L, i, t}-P_{L, i, t}^{D} \geq 0  \perp \rho_{5,i,t}^{max}\geq 0  ;\forall i \in I, \forall t \in T\label{de13}&
\end{align}
where $i(v)$ denotes the bus at which unit $v$ is located. Conditions (\ref{de1})–(\ref{de5}) are constraints of the dual problem of (\ref{e1})–(\ref{e6}) while (\ref{de6})–(\ref{de13}) are complementarity conditions related to the inequality constraints. 
\subsection{Optimality Conditions for Natural Gas Market Operation Model}
A set of necessary and sufficient conditions for a global optimum of (\ref{g1})–(\ref{g5}) is (\ref{g2})–(\ref{g5}) and:
\begin{align}
&C_{S,w}-\mu_{m(w),t}+\phi_{1,w,t}^{max}-\phi_{1,w,t}^{min}=0;\forall w \in W,\forall t \in T \label{dg1}
\end{align}
\begin{align}
&-C_{G}+\mu_{m,t}+\phi_{3,m,t}^{max}-\phi_{3,m,t}^{min}=0; \forall m \in M, \forall t \in T \label{dg2}
\end{align}
\begin{align}
& \sum_{n \in \mathbb{N}_m} (\mu_{m,t}-\mu_{n,t})+\sum_{n \in \mathbb{N}_m} (\phi_{2,m,n,t}^{max}-\phi_{2,n,m,t}^{max})  \notag\\
& -\sum_{n \in \mathbb{N}_m} (\phi_{2,m,n,t}^{min}-\phi_{2,n,m,t}^{min}) =0; \forall w \in W, \forall t \in T \label{dg3}\\
& F_{L,m, t}^{D} \geq 0 \perp \phi_{1,m,t}^{min} \geq 0; \forall m \in M,\forall t \in T \label{dg4}\\
& F_{L,m,t}- F_{L,m, t}^{D} \geq 0   \perp \phi_{1,m,t}^{max}\geq 0  ;\forall m \in M,\forall t \in T \label{dg5}\\
& F_{m,n, t}+F_{m,n,t}^{max}   \geq 0  \perp \phi_{2,m,n,t}^{min}\geq 0  ;\forall m \in M, n \in\mathbb{N}_m  ,\forall t \in T \label{dg6}\\
& F_{m,n,t}^{max}- F_{m,n, t} \geq 0  \perp \phi_{2,m,n,t}^{max}\geq 0  ;\forall m \in M, n \in\mathbb{N}_m ,\forall t \in T \label{dg7}\\
& F_{S,w, t}-F_{S,w,t}^{min} \geq 0    \perp \phi_{3,w,t}^{min}\geq 0  ;\forall w \in W,\forall t \in T \label{dg8}\\
& F_{S,w,t}^{max}- F_{S,w, t} \geq 0  \perp \phi_{3,w,t}^{max}\geq 0  ;\forall w \in W,\forall t \in T \label{dg9}
\end{align}
where $m(w)$ denotes the node at which gas source $w$ is located. Conditions (\ref{dg1})–(\ref{dg3}) are constraints of the dual problem of (\ref{g1})–(\ref{g5}) while (\ref{dg4})–(\ref{dg9}) are complementarity conditions related to the inequality constraints.
\subsection{Optimality Conditions for CEM Operation Model}
A set of necessary and sufficient conditions for a global optimum of (\ref{c2})–(\ref{c5}) is
\begin{align}
&-C^{N}+p_{co_2,t_c}+\nu_{1,o,t_c}^{max}-\nu_{1,o,t_c}^{min}=0;\forall o \in O, \forall t_c \in T_c \label{dc1}
\end{align}
\begin{align}
&C_{C,r}-p_{co_2,t_c}+\nu_{2,r,t_c}^{max}-\nu_{2,r,t_c}^{min}=0; \forall r \in R, \forall t_c \in T_c \label{dc2}\\
& Q_{L,o, t_c}^{D}\geq 0   \perp \nu_{1,o,t_c}^{min}\geq 0  ;\forall o \in O, \forall t_c \in T_c \label{dc3}\\
& Q_{L,o,t_c}- Q_{L,o,t_c}^{D} \geq 0   \perp \nu_{1,o,t_c}^{max}\geq 0  ;\forall o \in O, \forall t_c \in T_c \label{dc4}\\
& Q_{L,r, t_c} \geq 0    \perp \nu_{2,r,t_c}^{min}\geq 0  ;\forall r \in R, \forall t_c \in T_c \label{dc5}\\
& Q_{ L,r,t_c}^{max} -Q_{L,r, t_c} \geq 0  \perp \nu_{2,r,t_c}^{max}\geq 0  ;\forall r \in R, \forall t_c \in T_c. \label{dc6}
\end{align}
Conditions (\ref{dc1})–(\ref{dc2}) are constraints of the dual problem of (\ref{c2})–(\ref{c5}) while (\ref{dc3})–(\ref{dc6}) are complementarity conditions related to the inequality constraints.

Finally, taking these optimality conditions together, we search for equilibrium by solving the following optimization problem:
\begin{align}
\min 1
\end{align}
\begin{align}
&(\ref{c3})-(\ref{c5}),(\ref{e2})-(\ref{e7}),(\ref{g2})-(\ref{g5}),\notag\\
&(\ref{de1})-(\ref{de13}),(\ref{dg1})-(\ref{dg9}),(\ref{dc1})-(\ref{dc6})
\end{align}

Now the proposed model is also an MPEC problem. Note that the complementary-slackness conditions in KKTs can be linearized using the technique that is proposed in \cite{fortuny1981representation}. The form $a \perp b $ can be replaced $a \geq 0$, $b \geq 0$, $a \leq \phi M$ and $ab\leq (1-\phi) M$, where $\phi$ is a binary variable and $M$ is a sufficiently large positive
constant. Then the problem is transformed into a MILP problem. The problem is programmed using python-based open-source software package Pyomo \cite{hart2011pyomo} and solved using CPLEX.



\section{Case Study}

\begin{figure}[t]
\centering
\includegraphics[scale=0.8]{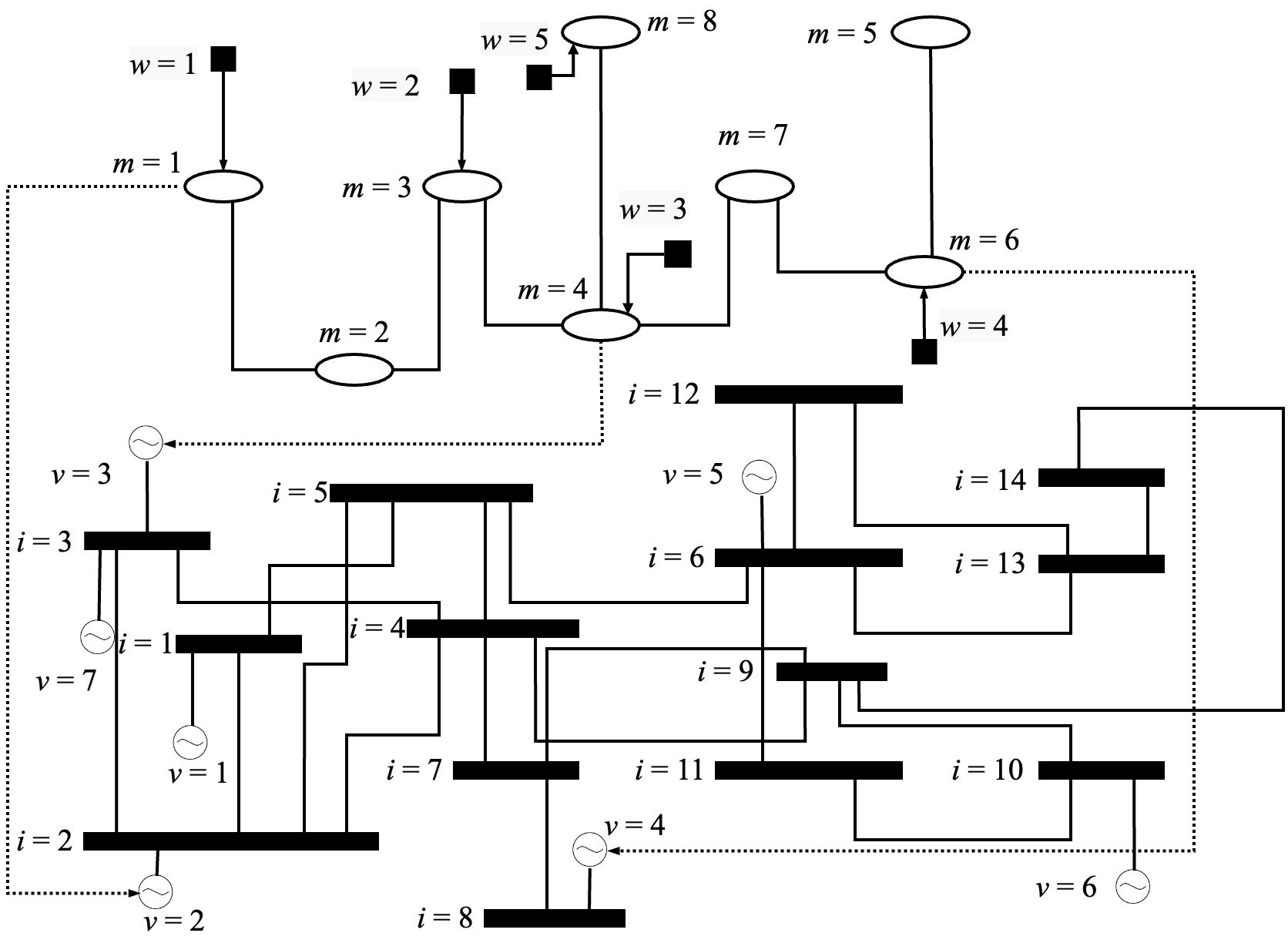}
\caption{One-line diagram of the 14-bus power system and 8-node natural gas system.}
\label{fig:1}
\end{figure}

In this section, we illustrate the performance of the proposed approach based on a regional energy system that consists of a 14-bus power system \cite{IEEE14bus}  and an 8-node natural gas system. Fig. \ref{fig:1} shows the topology of the system under study.

There are 7 generators in the system, including 6 dispatchable units and one wind generator. The parameters of the dispatchable generators are given in Table \ref{tab:1}. There are two coal-fired generators $G_1$ and $G_6$, which have low production costs and high carbon emission rates. The natural gas-fired generation units $G_2$, $G_3$, and $G_4$ are located at electric buses 3, 5, and 7 and connected to natural gas nodes 4, 4, and 3, respectively. Besides, there is one zero-carbon clean-fuel (e.g., hydrogen) generator $G_5$, which has limited capacity and high production costs. The wind power generator $G_7$ is located at bus 3. The electricity demand for each bus is given in \cite{github}. The parameters of the natural gas system are given in Table \ref{tab:3}. 

As for the CEM, we assume that there are 7 carbon allowance suppliers and 2 carbon allowance demands in this region. The regional emission cap is set to 225 tons and has been pre-allocated to the carbon allowance suppliers. The parameters of the CEM are given in Table \ref{tab:5}. 
\begin{table*}[!h]

\centering
\caption{Parameters of the dispatchable generators}
\label{tab:1}
\begin{tabular}{ccccccccc}
\toprule
Generator number &Type& $P_{min}$(MW) & $P_{max}$(MW) & $C_{G,v}$(\$/MW)& $C_{O,v}$(\$/MW)& Gas node & $\zeta_v$  (M$\rm m^3$/D/MW) &$\eta_v$(ton/MW) \\
\midrule
$G_1$ & coal-fired &0&80& 8.95&-&-&-&0.825 \\

$G_2$ & gas-fired &0&70& -&3.5&4&0.006&0.425\\

$G_3$ & gas-fired &0&60& -&1.5&4&0.007&0.435\\ 

$G_4$ & gas-fired &0&60& -&2.5&3&0.0065&0.435 \\

$G_5$ & clean-fuel-fired &0&30&21.90& -&-&-&0 \\

$G_6$ & coal-fired  &0&80&9.5& -&-&-&0.625\\
\bottomrule
\end{tabular}
\end{table*}

\begin{table}[h]

\centering
\caption{Parameters of the natural gas sources}
\label{tab:3}
\begin{tabular}{ccccc}
\toprule
Name & $F_{S,w}^{min}$ (M$\rm m^3$) &$F_{S,w}^{max}$ (M$\rm m^3$)&  Cost(\$/M$\rm m^3$) & Gas node\\
\midrule
$W_1$ & 0&1 &2090&1\\

$W_2$&  0&1.2& 2100&3\\

$W_3$ & 0&1.1& 2110&4\\

$W_4$ & 0&1.2&2200&6\\

$W_5$ & 0&0.9&2300&8\\
\bottomrule
\end{tabular}
\end{table}

\begin{table}[h]

\centering
\caption{Parameters of the CEM}
\label{tab:5}
\begin{tabular}{cccc}
\toprule
Name & Type &  Amount (ton) &  Cost (\$/ton) \\ 
\midrule
$S_1$ &Carbon supply&60 &12 \\

$S_2$& Carbon supply &50&15\\

$S_3$ &Carbon supply &40&16\\

$S_4$ &Carbon supply &30&18\\

$S_5$ &Carbon supply &20&20\\ 

$S_6$ &Carbon supply &15&25\\ 

$S_7$& Carbon supply &10&26\\

$CD_1$ &Carbon demand &20&-\\

$CD_2$ &Carbon demand &10&-\\
\bottomrule
\end{tabular}
\end{table}

\subsection{Effects of Electricity and Natural Gas Demand Change}

We start by evaluating the effects of electricity and natural gas demand growth/decline on the equilibrium of the three markets. Table \ref{tab:6} shows how different markets respond to the electricity load variation in terms of market clearing prices and carbon emissions in the proposed integrated energy system, respectively. Note that a \% denotes the deviation of the magnitude of the load from its baseline value. It should also be noted that since there are many nodes and buses in the system, we calculate their average LMPs in the following analysis.

As shown in Table \ref{tab:6}, with the rising electricity demand, the electricity, natural gas, and carbon allowance prices within the integrated energy system all increase. However, based on the generation output profile as shown in Table \ref{tab:genout}, we can observe that with load growth, the dispatch of the coal generator $G_1$, which is cheap but most emission-intensive, is gradually reduced in Cases 3-5. The outputs from the more expensive, but lower-emission gas-fired generator $G_2$ increase in Cases 3-5. Therefore, the electricity system is able to maintain the emission level. This observation shows that the proposed CEM mechanism provides an incentive in terms of encouraging low-emission generation units to participate in the electricity market in terms of increasing carbon prices while maintaining the current emission level. We can also observe in Cases 6 and 7 that when the load grows over a specific range, the system would run out of lower-emission generation capacities and the ISO has to rely on coal generation to satisfy the loads, which inevitably increases the carbon emission level despite the increased carbon allowance price. This observation confirms that in the face of electricity demand growth, due to long-term electrification of the economy or near-term extreme weather events caused by climate change, such as the recent heat waves in Europe and California in 2022, a sufficient zero-/low-emission generation capacity would play a key role in the realization of the deep decarbonization of electric power sector. 


\begin{table}[!h]

\centering
\caption{Electricity, natural gas, and carbon prices under varying electricity demands}
\label{tab:6}
\begin{tabular}{cccccc}
\toprule
Test & Electricity  & Electricity & Natural gas & Carbon &Carbon \\
case & demand & price & price & price &emission\\
 & growth & (\$/MWh)&(\$/M$\rm m^3$)&(\$/ton)&(ton)\\
\midrule
1& 0\% &23.8&2138.75&18& 134.38\\

2& 5\% &23.8&2138.75&18&145.07\\

3& 10\% &23.82&2138.75&18.02&150\\

4 & 15\% & 23.82&2140&18.02&150\\

5 & 20\% & 23.82 &2140&18.02&150\\

6 & 25\% &25.45 &2140&20&159.79\\

7 & 30\% &29.58&2140&25&170.48\\
\bottomrule
\end{tabular}
\end{table}

\begin{table}[!h]
\centering
\caption{Output of generators under varying electricity demands}
\label{tab:genout}
\centering

\begin{tabular}{p{0.5cm}<{\centering}p{0.7cm}<{\centering}p{0.7cm}<{\centering}p{0.7cm}<{\centering}p{0.7cm}<{\centering}p{0.7cm}<{\centering}p{0.7cm}<{\centering}}
\toprule
Test & $G_1$ & $G_2$ & $G_3$&  $G_4$ & $G_5$& $G_6$ \\
case & (MWh) & (MWh)  &(MWh)  &(MWh)  &(MWh) &(MWh)\\
\midrule
1& 39&0& 60.0& 60.0&30.0&80.0\\

2 &51.96&0&60.0& 60.0& 30.0&80.0\\

3 &50.54&14.36&60.00&60.0&30.0&80.0\\

4 &36.81&41.02&60.0&60.0&30.0&80.0\\

5 &23.02&67.78&60.0&60.0&30.0&80.0\\

6  & 33.75 &70 &60 &60 &30 &80 \\

7 & 46.7 &70.0&60.0&60.0&30.0&80.0\\
\bottomrule
\end{tabular}
\end{table}

Meanwhile, Table \ref{tab:7} depicts how different markets react to the varying load growth under the conventional cap-and-trade model. Compared with the results shown in Table \ref{tab:6}, we can observe that the carbon emissions are higher under the cap-and-trade model for all cases. This can be attributed to the carbon price being zero for Cases 1-6. In fact, the carbon price only becomes non-zero when the load growth reaches 30\% and the total carbon emission from the electricity generation reaches 195 tons, which makes the regional CEM reaches its cap of 225 tons. This observation is consistent with our previous assertion that the cap-and-trade model works only when the emission level raises close to the emission cap. In comparison, the proposed CEM model is capable of providing a consistent price signal to stimulate carbon mitigation even when the emission is low compared to the given cap. 

The above observation is supported by the generation output profile provided in Table \ref{tab:genout2}. Compared to the results presented in Table \ref{tab:genout}, it can be clearly observed that the ISO prioritizes the dispatch of the coal generator $G_1$ due to its cost advantages and the lack of a valid carbon price. On the other hand, the clean-fuel-fired generator $G_5$ is dispatched at last due to its high cost. This observation confirms our previous finding that the conventional cap-and-trade model may be unable to achieve the ambitious climate goal set for the power sector due to its sensitivity to the pre-determined cap. If such a cap is set too loose, it may not produce the appropriate price signal for Gencos and jeopardize the progress of deep decarbonization. On the other hand, if the cap is set too strict, it may constrain the normal operation of the electricity system and lead to events such as service interruptions and load shedding.

Furthermore, when we compare the electricity prices shown in Tables \ref{tab:6} and \ref{tab:7}, we can clearly observe that the electricity prices are higher in all cases under the proposed CEM model than in the cap-and-trade model. While this is expected and consistent with the estimation that maintaining or potentially lowering the emission level is likely to cause an increase in electricity and natural gas prices, the proposed market equilibrium model enables policymakers to quantitatively evaluate and balance the operations of different energy markets to assure that the increased energy prices will not negatively impact the welfare of society. In addition, the profits of the carbon suppliers in the market can also be re-distributed in an equitable way to mitigate the potential effects of increasing electricity prices, especially on marginalized and social-economically vulnerable communities.

\begin{table}[t]
\centering
\caption{Electricity, natural gas, and carbon prices under varying natural electricity demands with the cap-and-trade model}
\label{tab:7}
\begin{tabular}{cccccc}
\toprule
Test & Electricity  & Electricity & Natural gas & Carbon &Carbon\\
case & demand & price & price & price &emission\\
 & growth & (\$/MWh)&(\$/M$\rm m^3$)&(\$/ton)&(ton)\\
\midrule
1& 0\% &15.77&2138.75&0& 163.42\\

2& 5\% &16.16&2138.75&0&169.03\\

3& 10\% &16.16&2138.75&0&174.53\\

4 & 15\% & 16.16&2138.75&0&180.03\\

5 & 20\% & 16.16 &2140&0&185.54\\

6 & 25\% &16.16 &2140&0&191.04\\

7 & 30\% &21.91 &2140&13.52&195\\
\bottomrule
\end{tabular}
\end{table}

\begin{table}[t]
\centering
\caption{Output of generators under varying electricity demands with the cap-and-trade model}
\label{tab:genout2}
\centering

\begin{tabular}{p{0.5cm}<{\centering}p{0.7cm}<{\centering}p{0.7cm}<{\centering}p{0.7cm}<{\centering}p{0.7cm}<{\centering}p{0.7cm}<{\centering}p{0.7cm}<{\centering}}
\toprule
Test & $G_1$ &$G_2$ &$G_3$&  $G_4$ & $G_5$& $G_6$ \\
case & (MWh) & (MWh)  &(MWh)  &(MWh)  &(MWh) &(MWh)\\
\midrule
1& 80&0& 49& 60&0&80\\

2 &80&1.96&60& 60& 0&80 \\

3 &80&14.9&60&60&0&80\\

4 &80&27.83&60&60&0&80\\

5 &80&40.8&60&60&0&80\\

6  & 80 &53.75 &60 &60 &0 &80 \\
 
7 & 80 &63.06&60.0&60.0&3.64&80.0\\
\bottomrule
\end{tabular}
\end{table}

\subsection{Effects of Renewable Generation Adoption on the CEM}

In this subsection, we focus on analyzing how the continuous decarbonization of electricity generation can impact the operation of the integrated energy system. Specifically, we evaluate the effects of retrofitting existing generators in the electricity system to mitigate their carbon footprints. Different retrofitting strategies, such as modifying the specifications of the generation technologies or adding CCUS into the system, can effectively reduce the carbon emission rates of the generation units. However, deploying such strategies also increases the operation cost of the retrofitted units. Without loss of generality, in the following study, we assume that once a generation unit is retrofitted, its emission rate will be lowered to 0.1 ton/MW. We also assume that following retrofitting, the operation cost $C_{G,v}$ of $G_1$ is increased to 15 \$/MW, and the $C_{O,v}$ of $G_2$ and $G_3$ becomes 7 \$/MW, respectively.

Table \ref{tab:9} shows the effects of modifying different generators on the prices of different markets. The result suggests that modifying generators can help reduce carbon emissions in general. However, the extent of such a reduction can be significantly different for different retrofitting strategies. For instance, it can be observed that retrofitting the most carbon-intensive generator $G_1$ can lower the emission level to 92.37 tons, while retrofitting $G_2$ can only reduce the emission level to 107.03 tons. Meanwhile, retrofitting $G_1$ results in a slightly higher electricity price. We can also observe that retrofitting $G_1$ and $G_2$, or retrofitting $G_1$ to $G_3$ altogether, can give us the best overall performance in terms of electricity price and carbon emission. As retrofitting a generator usually requires a large capital investment and takes a period of time to complete, the proposed model would allow Gencos and policymakers to prioritize and enhance the sequence of retrofitting to lower its impacts on electricity markets while meeting the decarbonization objectives.

Another important observation we can make based on Table \ref{tab:9} is that following retrofitting, the electricity price can be lowered despite the increasing operation cost for the retrofitted generators. This observation further demonstrates the effectiveness of the proposed CEM model in terms of stimulating clean and emission-free generation in the power sector.


\begin{table}[t]
\centering
\caption{Electricity, natural gas, and carbon prices under different retrofitting strategies}
\label{tab:9}

\begin{tabular}{p{2.4cm}<{\centering}p{1.1cm}<{\centering}p{1.1cm}<{\centering}p{1.1cm}<{\centering}p{1.1cm}<{\centering}}
\toprule

Retrofitting strategy & Electricity price (\$/MWh)& Natural Gas price (\$/M$\rm m^3$)&  Carbon price (\$/ton)& Carbon emission (ton)\\
\midrule
None &23.8	&2140.0&	18&134.38\\

Modifying $G_1$& 22.73	&2138.75&	16&92.37\\

Modifying $G_2$& 22.15	&2140.0&	16&107.03\\

Modifying $G_3$& 23.54	&2138.75&	17.68&120 \\

Modifying $G_1$, $G_2$ & 21.91&	2140.0&	15.53&80\\

Modifying $G_2$, $G_3$ & 22.15 &	2140.0&	16&107.03\\

Modifying $G_1$, $G_3$ & 22.96&	2138.75&	16&92.18\\

Modifying $G_1$ - $G_3$ & 21.91&	2106.25&	15.53&80\\
\bottomrule 
\end{tabular}
\end{table}

\subsection{Impacts of CEM Specifications}

\begin{figure}[!ht]
\centering
\includegraphics[width=.9\columnwidth]{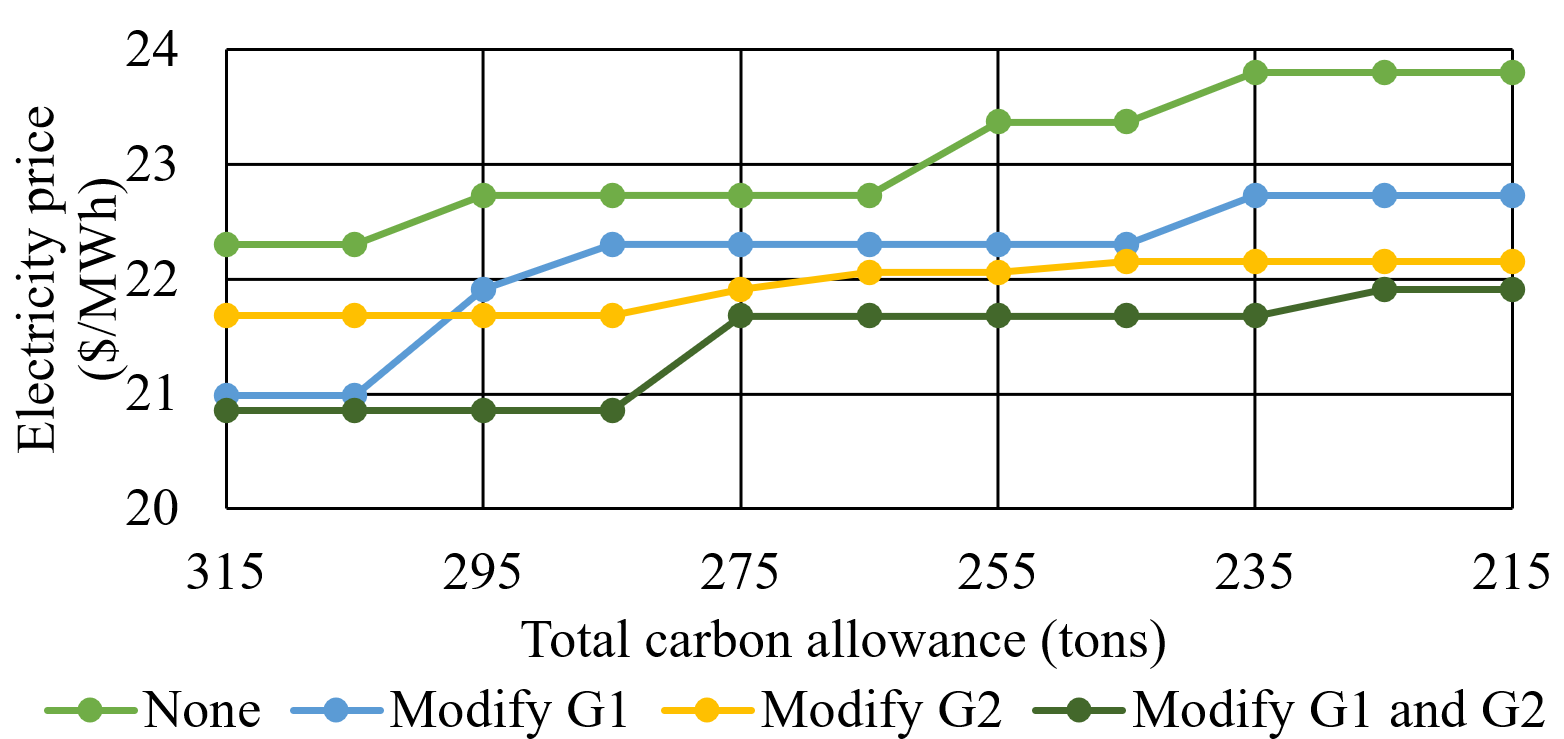}
\caption{Electricity price under different total carbon allowance}
\label{fig:5}
\end{figure}

\begin{figure}[!ht]
\centering
\includegraphics[width=.9\columnwidth]{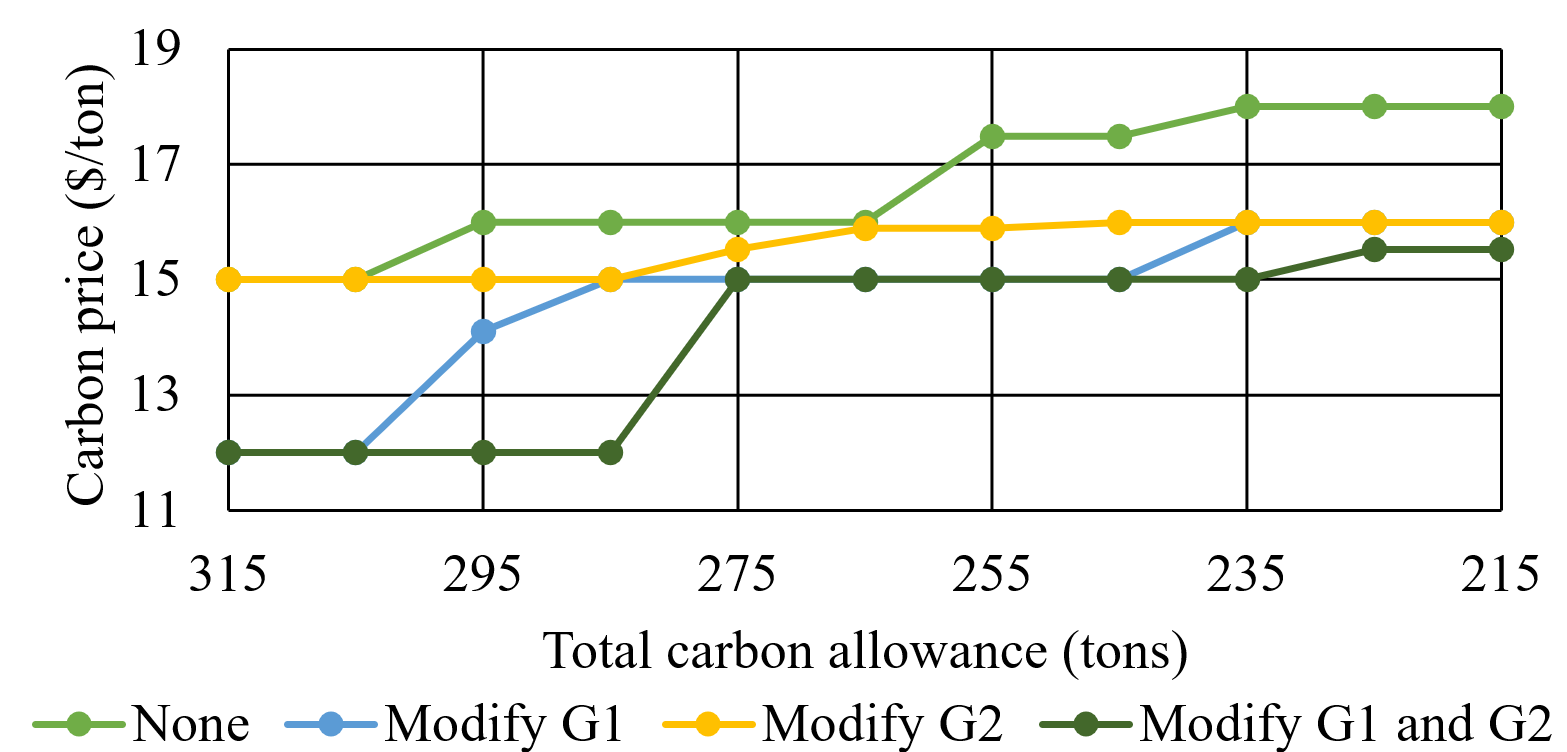}
\caption{Carbon price under different total carbon allowance}
\label{fig:6}
\end{figure}

In this subsection, we inspect how the specifications of the CEM can affect the operation of the integrated energy system. Specifically, Figs. \ref{fig:5} and \ref{fig:6} show how tightening the total carbon allowance impacts electricity and carbon prices, respectively. We can observe that achieving a more ambitious decarbonization objective (i.e., a lower total carbon allowance) would lead to rising electricity and carbon allowance prices. This observation is consistent with our previous analysis. Such a dilemma needs to be carefully evaluated by policymakers to strike a balance between power system economics and the progress of decarbonization. 

Additionally, we can observe that deploying more no-/low-emission technologies in the system can mitigate some of the negative effects mentioned above. Based on this observation, we recommend government-level initiatives, such as subsidies and financial assistance, to speed up the help steer the pathway of continuous and more in-depth decarbonization.

\begin{figure}[!ht]
\centering
\includegraphics[width=.9\columnwidth]{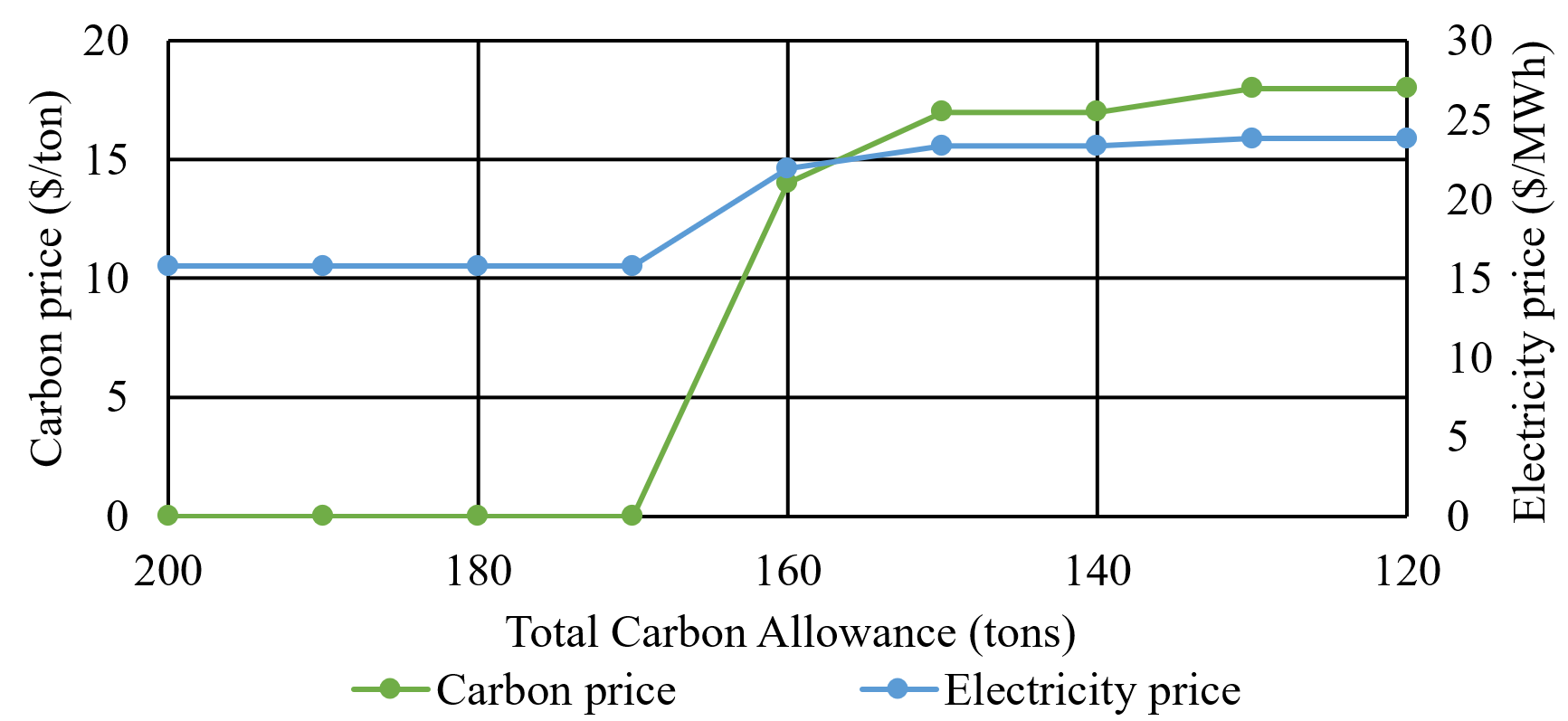}
\caption{Carbon price under a cap-and-trade CEM}
\label{fig:7}
\end{figure}

As a comparison, we evaluate the effects of tightening the total carbon allowance on the cap-and-trade model. The result of this analysis is shown in Fig. \ref{fig:7}. Similar to what we have observed in the previous analysis, Fig. \ref{fig:7} shows that the cap-and-trade does not work well until the emission cap of the emission trading system is set close to the current emission level, which is around 163 tons in our study. In other words, it relies heavily on the accurate section of the cap value to produce appropriate impacts on the Gencos and other electricity market participants. In comparison, the proposed model provides a more flexible and consistent way of regulating and controlling carbon prices to stimulate changes in the electricity system.
.

Finally, we study the effects of different market clearing time scalars on the carbon price and total system emission in the proposed model. The result of this analysis is given in Table \ref{tab:8}. We can clearly observe that different clearing time scalars lead to different carbon prices and total carbon emissions in the system. Specifically, a longer clearing time leads to a lower average carbon price and a higher average hourly carbon emission. This observation is consistent with our expectation as maintaining an hourly carbon balance can be straining for the CMO especially when the electricity demand is high and more carbon allowances are needed for a particular time period such as the peak hours. Meanwhile, if the CEM is cleared on a daily basis, the market participants will have more means to arrange and control their usage of carbon allowances. Therefore, if the goal of the CMO is to reduce the overall emission within a jurisdiction and make the CEM more active, it may want to select a shorter market clearing time. However, if a lower electricity price is more favorable for the jurisdiction, the CMO can increase the market clearing intervals.

\begin{table}[t]
\centering
\caption{Effects of different CEM clearing times}
\label{tab:8}

\begin{tabular}{p{2.4cm}<{\centering}p{2.8cm}<{\centering}p{2.8cm}<{\centering}}
\toprule
Time scalar & Average carbon emission per hour (ton/h)& Average carbon price (\$/ton)\\
\midrule 
per-hour&134.38&18.0\\

per-3-hour& 146.88&16.0\\
 
per-12-hour & 151.31 & 15.0\\
 
per-24-hour & 157.54 & 15.0\\
\bottomrule 

\end{tabular}
\end{table}

\section{Conclusion}

This paper develops an operational-equilibrium model for the integrated operation of the electricity, natural gas, and carbon-emission markets. First, we explore the role of a regional carbon operator, which operates and manages the centralized carbon allowance trading in a jurisdictional CEM model. We then develop the model of an integrated regional energy system where the operations of electricity, natural gas, and carbon emission systems are all interconnected. To compute their operational equilibrium, we replace the model with their optimality conditions and transform the resulting MPEC problem model into a MILP problem, which can be solved by CPLEX. Simulation results presented in the case studies show the advantages of the proposed CEM model compared to the conventional cap-and-trade mechanism in terms of flexibility, consistency, and most importantly, the effectiveness of emission mitigation. We have also shown that the proposed equilibrium model allows policymakers and energy market participants to balance the near-term social welfare and the long-term, strategic climate objectives based on jurisdictional characteristics such as emission intensity, average household income, load profile patterns, as well as energy efficiency.

\bibliographystyle{IEEEtran}
\bibliography{main.bib}

\end{document}